\documentclass[twocolumn]{aastex7}
\usepackage{kotex}
\usepackage{multirow}
\usepackage{gensymb}
\usepackage{amsmath}
\usepackage{graphicx}
\usepackage{xcolor}
\usepackage[T1]{fontenc} 
\usepackage{enumitem}
\usepackage{comment}
\usepackage{appendix}
\shorttitle{X-ray Reverberation in NGC~4151}
\shortauthors{J. M. Miller et al.}

\begin{document}

\title{Reverberation in the Narrow Fe~K${\alpha}$ Line in the Seyfert Galaxy NGC~4151 with XRISM}

\author[orcid=0000-0003-2869-7682]{Jon M. Miller}
\affiliation{Department of Astronomy, University of Michigan, 1085 South University Avenue, Ann Arbor MI 48109, USA}
\email[show]{jonmm@umich.edu}

\author[0000-0002-0572-9613]{Abderahmen Zoghbi}
\affiliation{Department of Astronomy, The University of Maryland, College Park, MD 20742, USA}
\affiliation{HEASARC, Code 6601, NASA/GSFC, Greenbelt, MD 20771, USA}
\affiliation{CRESST II, NASA Goddard Space Flight Center, Greenbelt, MD 20771, USA}
\email{azoghbi@umd.edu}

\author[0000-0002-7129-4654]{Xin Xiang}
\affiliation{Department of Astronomy, University of Michigan, 1085 South University Avenue, Ann Arbor MI 48109, USA}
\email{xinxiang@umich.edu}

\author[orcid=0000-0002-3687-6552]{Doyee Byun (변도의)}
\affiliation{Department of Astronomy, University of Michigan, 1085 South University Avenue, Ann Arbor MI 48109, USA}
\email{doyeeb@umich.edu}

\author[0000-0002-4992-4664]{Missagh Mehdipour}
\affiliation{Department of Astronomy, University of Michigan, 1085 South University Avenue, Ann Arbor MI 48109, USA}
\email{missagh@umich.edu}

\author[0000-0001-9911-7038]{Liyi Gu}
\affiliation{SRON Space Research Organization Netherlands, Niels Bohrweg 4, 2333CA Leiden, The Netherlands} 
\email{l.gu@sron.nl}

\author[0000-0001-9735-4873]{Ehud Behar}
\affiliation{Department of Physics, Technion, Technion City, Haifa 3200003, Israel}
\email{behar@physics.technion.ac.il}

\author[0000-0003-2663-1954]{Laura Brenneman}
\affiliation{Center for Astrophysics | Harvard-Smithsonian, MA 02138, USA}
\email{lbrenneman@cfa.harvard.edu}

\author[0000-0001-8470-749X]{Elisa Costantini}
\affiliation{SRON Space Research Organization Netherlands, Niels Bohrweg 4, 2333CA Leiden, The Netherlands} 
\affiliation{Anton Pannekoek Institute for Astronomy, University of Amsterdam, Science Park 904, NL-1098 XH Amsterdam, The Netherlands}
\email{E.Costantini@sron.nl}

\author[0009-0006-4968-7108]{Luigi Gallo}
\affiliation{Department of Astronomy and Physics, Saint Mary's University, Nova Scotia B3H 3C3, Canada}
\email{lgallo@ap.smu.ca}

\author[0000-0002-1094-3147]{Matteo Guainazzi}
\affiliation{European Space Agency (ESA), European Space Research and Technology Centre (ESTEC), 2200 AG Noordwijk, The Netherlands} 
\email{Matteo.Guainazzi@esa.int}

\author[0000-0003-4511-8427]{Peter Kosec}
\affiliation{Center for Astrophysics | Harvard-Smithsonian, MA 02138, USA}
\email{peter.kosec@cfa.harvard.edu}

\author[0000-0002-2933-048X]{Takeo Minezaki}
\affiliation{Institute of Astronomy, School of Science, University of Tokyo, 2-21-1 Osawa, Mitaka, Tokyo 181-0015, Japan}
\email{minezaki@ioa.s.u-tokyo.ac.jp}

\author[0000-0002-8108-9179]{Stephane Paltani}
\affiliation{Department of Astronomy, University of Geneva, Versoix CH-1290, Switzerland} 
\email{Stephane.Paltani@unige.ch }

\begin{abstract}
Emission lines that ``echo'' variations in the ionizing flux produced close to black holes are powerful probes of the central engine.  In the Seyfert-1.5 galaxy NGC 4151, high-resolution X-ray spectra and time lags in low-resolution X-ray data suggest that part of the narrow Fe~K$_{\alpha}$ line originates close to the optical broad line region (BLR).  We report on a sequence of nine XRISM observations of NGC~4151, obtained every other day in 2024.  Swift monitoring was undertaken to sample the driving flux before, during, and after the XRISM sequence.  Using suitable line kernels, we measure a mean BLR component width of $\sigma = 5.36\pm 0.48\times 10^{3}~{\rm km}~{\rm s}^{-1}$.  Modeling the Swift continuum and XRISM line flux trends gives a lag of $\tau = 3.5^{+2.8}_{-1.7}$~days ($r=3.6^{+3.0}_{-1.7}\times 10^{3}~GM/c^{2}$ for $M_{BH} = 1.7\times 10^{7}~M_{\odot}$), significant at the $2\sigma$ level via Monte Carlo simulations, and consistent with prior measurements and direct spectral fits.  This lag implies a black hole mass of $M_{BH}/f_{X} = 2.0^{+1.4}_{-1.0}\times 10^{7}~M_{\odot}$, where $f_{X}$ is a geometrical factor.  A standard optical value for this factor gives a mass that is nominally higher than typical H$\beta$ mass estimates, but formally consistent.  Our results suggest that XRISM can measure lags and black hole masses in both unobscured and obscured AGN.
\end{abstract}


\section{Introduction}

NGC~4151 is an especially proximal ($z=0.0033$, \citealt{baer-way2024}) and well-studied Seyfert-1.5 active galactic nucleus (AGN).  It was one of the first AGN wherein  broad optical lines were found to lag continuum emission (e.g., \citealt{gaskell1986}), providing an early measure of length scales within the central engine and a window on the nature of the broad line region (BLR).  This technique is now used to constrain the mass of black holes in Seyfert galaxies and quasars; in NGC~4151, typical H$\beta$ lag measurements suggest a black hole mass consistent with $M_{BH} = 3.4\pm 0.4\times 10^{7}~M_{\odot}$ \citep{bentz2015}.  More recent work incorporating kinematics within the BLR gives a mass of $M_{BH} = 1.66^{+0.48}_{-0.34}~\times 10^{7}~M_{\odot}$ \citep{bentz2022}.

The proximity of NGC~4151 has helped to reveal the structure of the central engine in AGN in other important ways.  Radio observations with the VLBI, for instance, place a limit on the size of the radio continuum source -- potentially a compact jet -- of just 0.035~pc \citep{ulvestad2005}.  
Recent X-ray spectra find outflows with velocities of $v\simeq 0.15c$ \citep{xiang2025} -- faster than the low $v \leq 0.05c$ velocities inferred in the radio jet \citep{ulvestad2005} -- raising questions about the relationship between fast winds and jets. Dust reverberation lags in NGC~4151 suggest that the inner wall of the torus lies  just 0.033~pc ($r \simeq 1.9\times 10^{4}~GM/c^{2}$) from the black hole \citep{lyu2021}.  This value is broadly confirmed via IR interferometry \citep{gravity2023}.  

Again owing to its proximity, NGC~4151 is the brightest Seyfert-1 AGN above roughly 4~keV, with the highest flux in its narrow Fe~K$_{\alpha}$ emission line.  Whereas Chandra/HETG spectroscopy of fainter AGN did not conclusively associate the narrow Fe~K$_{\alpha}$ line with the optical BLR (for a review, see \citealt {shu2010}), direct fits to the Fe~K$_{\alpha}$ line in the summed Chandra spectrum of NGC~4151 suggested an emission radius of $r\simeq 1000~GM/c^{2}$ \citep{miller2018}, and variability studies with XMM-Newton and Suzaku revealed the first evidence of a BLR-like time lag in this line ($\tau = 3.3^{+1.8}_{-1.7}$~days, \citealt{zoghbi2019}).  At CCD resolution and even gratings resolution, however, different contributions to the narrow Fe~K$_{\alpha}$ line cannot be discerned, possibly leading to distortions in both direct fits and lag signals.

The Resolve calorimeter \citep{ishisaki2022} aboard XRISM \citep{tashiro2025} has rapidly improved our understanding of X-ray spectra from AGN, and how different features map the geometry of the accretion flow.  It is now evident that the ``narrow'' Fe~K$\alpha$ emission line in not a monolith, but rather a composition of flux from the optical BLR and the ``torus'' (e.g., \citealt{miller2024}, \citealt{miller2025}, \citealt{mehdipour2025}, \citealt{kammoun2025}, \citealt{li2026}, \citealt{miller2026}, \citealt{juranova2026}).  The same work highlights the need for physical models that include up-to-date atomic data, including Fe~K$_{\alpha,1}$ and K$_{\alpha,2}$ components and scattering effects (see, e.g., \citealt{murphy2009}, \citealt{tanimoto2019}).  

Clearly associating reverberation in the narrow Fe~K$_{\alpha}$ line complex with its BLR-like flux component would ground the assumptions made in direct spectral fits, place X-ray reverberation signals on firmer ground, and potentially build towards a new means of measuring black hole masses.  Given its high flux and compelling early results from XRISM, NGC~4151 is the most natural candidate for an initial trial.  For this reason, we proposed a series of nine observations of NGC~4151 during XRISM Cycle 1.  The observations were taken every other day for a span of 18 days, based on a mean H$\beta$ lags of $\tau \sim 5.5$~days \citep{bentz2022}.  Monitoring observations with the Neil Gehrels Swift observatory were made at a higher cadence, in order to trace the driving continuum flux in NGC~4151. 

This paper is organized into four sections.  The observations and data reduction procedures are described in Section 2.  Our analyses of the XRISM data, Swift data, and time lags are reported in Section 3. Finally, in Section 4, we discuss our results, compare them to past work, and comment on their possible implications for for the near- and long-term future of X-ray studies of AGN.

\section{Observations and Data Reduction}

The reduction and analysis steps presented in this work made use of the tools in HEASOFT version 6.34, and concurrent calibration database (CALDB) files.  The Swift/XRT spectra were analyzed using XSPEC version 12.14.1 \citep{arnaud1996}.  This choice was made for simplicity and speed, given the large number of spectra with modest sensitivity and resolution.  In contrast, the XRISM spectra were analyzed using SPEX version 3.08.01 \citep{kaastra1996}, in order to utilize the same models and techniques employed in recent fits to Resolve spectra of NGC~4151 (\citealt{miller2024}, \citealt{miller2026}).  All of the errors reported in this work are $1\sigma$ uncertainties, unless otherwise noted.

\subsection{XRISM}

We utilized a sequence of nine XRISM observations of NGC~4151 obtained at an every-other-day cadence during Cycle 1.  The Resolve \citep{ishisaki2022} spectra were initially discussed in \cite{miller2026}, as part of an analysis of the time-averaged 0.9~Ms spectrum.  These spectra are also the subject of a connected paper that treats variability in the wind absorption spectrum of NGC~4151 (Xiang et al.\ 2026, submitted).  Our analysis differs from prior XRISM treatments of NGC~4151 in that it also makes use of simultaneous data obtained with the Xtend CCD camera \citep{hayashida2018}.  Although its CCDs only deliver moderate spectral resolution, Xtend has a higher collecting area and -- especially in joint spectral fits we have undertaken -- can help to constrain variability in line components once the proper model is determined with Resolve.   The observation identification numbers (ObsIDs) and the MJD of the {\em midpoint} of each Resolve exposure are listed in Table 1 (the midpoint of each observation was used in the time lag analysis; see below).

In each observation, Resolve was run in ``PX\_NORMAL'' mode.  Following the recipe in version 2.3 of the XRISM Quick Start Guide, we filtered the data to: (1) exclude times close to passage through the South Atlantic Anomaly (SAA), (2)  exclude pixels 11, 12, and 27 (the calibration pixel and those that sometimes read anomalously), and (3) only include high-resolution primary events (Hp).  Spectra were then generated using \texttt{xselect}, while calibrated response files were generated using \texttt{rslmkrmf} and \texttt{xaarfgen}.  We elected to create a ``Large'' redistribution matrix file.  

Owing to the fact that the X-ray flux from NGC~4151 can reach $F\geq 1\times 10^{-10}~{\rm erg}~{\rm cm}^{-2}~{\rm s}^{-1}$, or higher, Xtend was run in ``1/8 Window'' mode to prevent photon pile-up.  The reduction of the Xtend data followed the corresponding recipe within the XRISM Quick Start Guide. The data were filtered to: (1) exclude times close to the SAA, (2) exclude events from the onboard gain calibration sources, and (3) remove anomalous and flickering pixels.  In each observation, we also checked the attitude stability of the spacecraft; no anomalies were found.  Source and background spectra were then extracted using \texttt{xselect}.  Response files were generated using the tools \texttt{xtdrmf} and \texttt{xaarfgen}. 

\subsection{Swift}

In order to track the evolution of the X-ray flux in NGC~4151 before, during, and after the sequence of XRISM observations, we obtained a series of monitoring observations at a cadence of once or twice per day, for a period of two months.  The Swift/XRT provides imaging spectroscopy across the 0.3--10.0~keV band.  All of the XRT exposures in our sequence were obtained in ``Photon Counting'' mode to enable robust characterizations of the background.  For expedience and uniformity, we downloaded calibrated source and background spectra and responses from the Swift XRT Product Generator online facility, hosted by Leicester University \citep{evans2009}.  This tool provides a rapid and uniform reduction of data, and corrects spectra and responses for bad columns and pixels via exposure maps.

\section{Analysis and Results}

\subsection{XRISM Spectra}

The Resolve and Xtend spectra from individual observations were fit jointly using SPEX version 3.08.01 \citep{kaastra1996}, minimizing a Cash statistic \citep{cash1979}.  Prior to fitting, the spectra were binned using the ``optimal'' binning algorithm \citep{kaastra2016}, and subsequently by a factor of 3.0 to further boost their sensitivity.  

In preliminary joint fits to each of the nine observations, we found excellent agreement between the Resolve and Xtend continuum flux levels over the 4.4--8.40~keV band.  We adopted this fitting band, and proceeded to fit the spectra without introducing a flux normalization between Resolve and Xtend.  The narrow fitting band enables a robust characterization of the Fe~K$_{\alpha}$ emission line complex.  

In developing a spectral model, we emphasized simplicity to avoid biases that might follow from complex continua or wind models.  We modeled the spectra using a streamlined version of the model applied to the time-averaged Resolve spectrum of NGC~4151 \citep{miller2026}:\\

\noindent$\bullet$ First, we assumed a continuum consisting of a simple power-law model attenuated at low and high energies using ``etau'' components.  The free parameters included the power-law index and flux normalization.\\

\noindent$\bullet$ Second, we modeled cool partial-covering absorption using the ``hot'' component, allowing its temperature to vary within the range found in the time-averaged Resolve spectrum ($kT = 5.5^{+0.4}_{-0.3}$~eV; \citealt{miller2026}).  Free parameters initially included the column density, gas temperature, and covering fraction.  However, we found that the temperature was poorly constrained and the covering factor always included unity, so we used the best-fit temperature from \cite{miller2026} and set $f_{cov} = 1$ in all fits (leaving only the column density free to vary).\\  

\noindent$\bullet$ Third, we decomposed the narrow Fe~K$_{\alpha}$ emission line into torus and BLR components, modeling each with Gaussian-broadened ``xclumpy'' line kernels \citep{tanimoto2019}.  The ``xclumpy'' kernels include key atomic structure that is evident at calorimeter resolution (e.g., K$_{\alpha,1}$, K$_{\alpha,2}$, and K$_{\beta}$ lines in appropriate ratios), and a smooth red wing due to down-scattering in gas and dust with bound electrons.  Key ``xclumpy' parameters include the gas column density (fixed to the best-fit values measured in each component in the time-averaged spectrum; \citealt{miller2026}), the angular width and inclination of the gas region (both fixed at 45~degrees, as these parameters are largely unimportant for Gaussian broadening), the power-law index of the continuum (linked to the value of the same continuum parameter), the value of the power-law cut-off energy (fixed to the model limit of 370~keV), and the flux normalization.  The broadening was accomplished by appling a ``vgau'' component.  For the torus and BLR line components, then, the free model parameters included the line flux normalization and Gaussian broadening width ($\sigma$). \\

The results of fits with this model are shown in Figures 1 and 2, and detailed in Table 1.  The values of parameters such as the power-law index and absorption column density are reported, but should merely be regarded as shape parameters in this analysis.  Especially when only fit over a limited band, they do not reflect values that truly characterize the source.  Prior fits to individual XRISM observations across a broad pass band measured power-law indices broadly consistent with $\Gamma = 1.7$, and those fits are likely a better reflection of the character of the true driving flux.  The line component parameters were error scanned on a component-by-component basis, and then again with all parameters in order to reveal any degeneracies.  The line parameters were found to be independent of other parameter values, likely because Resolve effectively isolates BLR line flux from neighboring line and continuum flux.  All of the emission line parameters are extremely well determined, with typical errors of 10\% on line width and 5--6\% on line flux.  The fits are not formally acceptable because no attempt was made to fit the complex narrow absorption line spectrum.  However, the strongest wind lines in this band -- due to Fe XX-XXVI -- are associated with distant warm absorbers that vary only modestly \citep{xiang2025}, so they are unlikely to distort the BLR emission line from observation to observation.

Figure 1 shows the fit to each spectrum over the 4.4--8.4~keV band to verify the total model over the fitting band. Figure 2 shows the spectra over the 5.8--7.2~keV band to better illustrate the details of the line modeling.  To assess the statistical significance of the line variations, we calculated the mean line flux normalization and width of the BLR component, froze this value in fits to each spectrum, noted the change in the Cash statistic for two degrees of freedom, and calculated the resulting change an Akaike Information Criterion \citep{akaike1974} using the form implemented by \cite{xiang2025}.  Briefly, models are equally good when $\Delta{\rm AIC} > -2$, a more sophisticated model (in this case, a variable line flux rather than a constant line flux) is preferred for $-10 < \Delta{\rm AIC} < -2$, and a more sophisticated model is strongly preferred for  
$\Delta{\rm AIC} < -10$.  A variable line is most strongly indicated by observations -080 and -090, for which $\Delta{\rm AIC} = -14.9$ and $\Delta{\rm AIC} = -16.9$, respectively.  Variability is also preferred for observations -040 and -060; for other observations, variations in the torus component are better able to compensate for a static BLR component flux.

It is apparent that the width of the BLR line component varies more than its peak intensity.  In some AGN, enhanced continuum flux leads to more distant radii being illuminated, creating a narrower line profile in response; this is known as ``breathing'' (see, e.g., \citealt{wang2020}).   Some ``breathing'' behavior could be driven by heightened wind production when the continuum flux is higher.  In other AGN, ``anti-breathing'' is observed; this could be the result of a changing coronal geometry, or due to enhanced wind production in response to enhanced continuum flux.   This is discussed in more detail in Section 4.

\subsection{Swift Monitoring}

Prior to fitting within XSPEC, the Swift/XRT spectra were grouped using the ``optimal'' binning algorithm of \cite{kaastra2016}.  Although the Swift/XRT is sensitive over the 0.3--10.0~keV band, we narrowed the fitting range to the 0.5--10.0~keV band to avoid calibration uncertainties close to the edge of the pass band, and diffuse emission from the extended ``narrow line region'' in NGC~4151 (e.g., \citealt{kraemer2020}).  The XRT spectra were each fit with a simple model consisting of a simple power-law continuum, modified by partial covering via a ``tbpcf'' component \citep{wilms2000}.  The fits minimized a Cash statistic \citep{cash1979}.  For simplicity, the power-law index was fixed at a value of $\Gamma = 1.7$, consistent with typical values in Seyfert galaxies \citep{nandra2007} and prior fits to XRISM data on a broad pass band (\citealt{miller2024}, \citealt{xiang2025}).  The model then measured the power-law flux normalization, partial covering column density, and partial covering factor ($0\leq f_{cov} \leq 1$).  Given the modest sensitivity of the spectra, this model achieved acceptable fits to the data.  For each spectrum, we calculated the measured (absorbed) flux, the unabsorbed flux (removing the partial covering column), and the unabsorbed flux in the XRISM/Resolve band (2.2--17.4~keV) for later correlation.

XRISM/Resolve spectra indicate that the total obscuring column in NGC~4151 is very complex, composed of several highly ionized wind components as well as low-ionization absorption (\citealt{miller2024}, \citealt{xiang2025}).  Moreover, it is likely that the dominant column at low energy is from {\em low-ionization} gas \citep{miller2026}.  Our simplified model approximates the sum of this opacity as single column of {\em neutral} gas.  This approach is not formally correct, but it is far superior to simply using the XRT count rate or an absorbed flux.  Importantly, it provides a characterization that is suitable for tracking the ionizing radiation that stimulated the emission line.  Table 2 lists the start times and inferred flux levels of the (unabsorbed) driving continuum based on these fits.

Figure 3 shows two examples of Swift monitoring spectra.  These illustrate the strong fluctuations in the internal column that serve to shape the spectrum of NGC~4151, and the need to account for absorption before assessing the flux from the central engine.  In the limit of a fixed power-law index, the Swift/XRT spectra enable a reliable trace of changes in the column density and flux.  We note that the absolute flux is not crucial to the lag analysis.  The extension of the Swift monitoring to periods long before and after the XRISM sequence, and the density of the Swift monitoring, are essential factors in conducting the lag analysis.  (In a future paper, we will present the obscuration history and obscuration power spectrum of NGC~4151 across two decades of Swift monitoring.)

\subsection{Lag Analysis}

Figure 4 shows the light curves of the unabsorbed X-ray flux from NGC~4151 inferred using fits to the Swift/XRT monitoring data (projected to the full XRISM band), and the flux of the Fe~K$_{\alpha}$ BLR line component inferred in joint fits to the Resolve and Xtend spectra.  The continuum flux varies by as much as a factor of $\sim2$, but shows 20--25\% variations closer in time to the XRISM sequence.  The BLR component flux is also variable at the 20--25\% level.  This already suggests that the BLR might reprocess ionizing flux fairly efficiently.  Moreover, the level of variability observed in the BLR component is consistent with expectations based on earlier indications of reverberation in the Fe~K$_{\alpha}$ emission line \citep{zoghbi2019}.  Visually comparing the two trends, it is clear that they are not an exact match, and that the Fe~K$_{\alpha}$ line flux may lag the continuum flux.

To assess whether or not the BLR component flux lags the continuum, we primarily utilized pyCCF \citep{sun2018} and made additional checks using Javelin \citep{zu2013}.  The power of pyCCF is that it is a simple interpolated cross-correlation function (ICCF) that makes no assumption about the intrinsic nature of the variability process or light curves.  In contrast, the default use of Javelin assumes that the driving light curve is a damped random walk process, and that the responding light curve is a smoothed, scaled, and lagging version of the driving light curve.  Javelin can also be run with a Matern kernel, a power-exponential kernel, or a Kepler-exponential kernel (most appropriate for quasi-periodic signals).

Given the limitations of our data -- just nine XRISM observations -- we initially searched for lags within a range of $-5 \leq \tau \leq 10$ days.  A negative lag is not physical -- the production of an Fe~K$\alpha$ line requires hard, ionizing radiation from the central engine -- but it is important to search this regime for strong secondary peaks that could arise from phase-wrapping or other issues.  We ran pyCCF using flux randomization and random subset sampling, with 10,000 Monte Carlo simulations.  Results were largely insensitive to the specific values used for the granularity of the interpolation grid (we selected 0.7~days), the threshold for considering an ICCF as significant (we selected 0.5), and the number of lag bins (we selected a value of 40).  Over this range and with these parameters, pyCCF returns a centroid lag of $\tau = 2.3^{+3.6}_{-2.0}$~days.  Ten percent of simulations find a negative lag but the distribution is a smooth continuation from a positive lag, not an anomalous secondary peak.  Over the same range, Javelin returns a lag of $\tau = 3.6^{+5.9}_{-2.1}$~days and a secondary maxima at negative values is also rejected. 

We then proceeded to examine lags over the physical range of $0 \leq \tau \leq 10$~days.  PyCCF measures a centroid lag of $\tau = 3.5^{+2.8}_{-1.7}$ and an ICCF correlation of $r = 0.47$.  This signals a $2\sigma$ detection of a lag between the ionizing X-ray continuum and the BLR component of the narrow Fe~K$\alpha$ line in NGC~4151.  Running Javelin in its default mode over this range results in a lag of $\tau = 3.5^{+2.1}_{-1.4}$~days.  The centroid lag probability distributions are shown in Figure 5.  The driving and responding light curves are shown again in Figure 6, before and after correcting for the lag.

A number of basic checks suggest that these results are robust.  Only running pyCCF with flux randomization but without random subset sampling, we obtain a consistent lag value of $\tau = 2.8^{+3.5}_{-1.0}$ days.  Eliminating one XRISM measurement from the data in sequence and running pyCCF with full flux randomization and random subset sampling an nine additional times, formally consistent lags are again obtained in each case.  Running Javelin with each of its alternative kernels also returns formally consistent lags.

Our Fe~K$_{\alpha}$ lag value is nominally shorter than the mean H$\beta$ lag of $\tau = 5.46^{+0.87}_{-0.78}$ days reported by \cite{bentz2022}, potentially suggesting that the Fe~K$_{\alpha}$ line may originate within an inner extension to the optical BLR though the two are formally consistent.  Some UV lines nominally show even shorter lags than the X-ray Fe~K$_{\alpha}$ line (e.g., C III] and He II, \citealt{mezroth2006}; C IV and Si IV, Byun et al.\ 2026 subm.), though they are also formally consistent within errors.  This suggests that dust, cold gas, and ionized gas may all exist at similar radii.  However, these components might be vertically stratified, with denser, cold, dusty gas emitting Fe~K$_{\alpha}$ lines close to the disk, and ionized UV lines emitted at greater scale heights.  (We note that some optical monitoring periods return short, potentially anomalous H$\beta$ lags of, e.g., $\tau = 3.25^{+1.40}_{-0.72}$~days; \citealt{feng2024}).

\subsection{Black Hole Mass Estimation}

In reverberation mapping studies, the mass of the black hole is given by ${M}_{BH} = f c\tau v^{2}/{G}$. Here, $\tau$ is the measured delay time between the ionizing continuum flux and responding line flux, $v$ is the velocity width of the line, and $f$ is a geometric scaling factor derived by comparing mass constraints from reverberation to those from stellar velocities in nearby galaxies.  The most robust framing of our mass constraint is then $M_{BH}/f_{X} = 2.0^{+1.4}_{-1.0} \times 10^{7}~M_{\odot}$.

The specific value of the black hole mass that is derived via reverberation (in any band) depends upon the correction factor, $f$.  In a study of numerous AGN, \cite{grier2013} derive a value of $f=4.31\pm 1.05$.  In work focused on NGC~4151 that incorporates kinematics within the broad line region, \cite{bentz2022} derive a smaller value of $f = 1.8^{+0.6}_{-0.4}$.  In order to match their mass of $M_{BH} = 1.66^{+0.48}_{-0.34}\times 10^{7}~M_{\odot}$, an X-ray correction factor of $f_{X} = 0.83$ would be required.  This is within about $2\sigma$ of the optical value and may indicate formal consistency.  We again note that specific values of $f$ can vary with the state of a given AGN, perhaps owing to changes within the BLR.

\section{Discussion}
We have obtained a series of nine XRISM observations of NGC~4151 at an every-other-day cadence.  This sequence was designed to oversample the H$\beta$ lag by a factor of three, for a string of three lag times.  Joint fits to the Resolve and Xtend spectra in the vicinity of the narrow Fe~K$_{\alpha}$ line easily separate the line flux in to narrow and broad components from the torus and BLR, respectively.  The flux of the BLR component is found to be variable at the 20--25\% level.  Using pyCCF to measure lags between continuum variations measured in a high-cadence Swift X-ray light curve and the BLR line component flux, we find $\tau = 3.5^{+2.8}_{-1.7}$~days, which closely corresponds to the lag obtained using CCD data \citep{zoghbi2019}.  The inferred Fe~K$_{\alpha}$ lag is nominally shorter than typical H$\beta$ lags but comparable to values found in some UV lines \citep{mezroth2006}.  The lag that we have measured implies a black hole mass of $M/f_{X} = 2.0^{+1.4}_{-1.0}\times 10^{7}~M_{\odot}$ in NGC 4151, where $f_{X}$ is an unknown geometric factor.  In this section, we compare our results to prior work, note some of its strengths and shortcomings, and discuss potential implications and future research directions.

Using a combination of CCD spectrometer data from Suzaku and XMM-Newton, \cite{zoghbi2019} derived an Fe~K$_{\alpha}$ lag of $\tau = 3.3^{+1.8}_{-0.7}$~days.  This is formally consistent with our value, likely indicating that Fe~K$_{\alpha}$ line reverberation in NGC 4151 is robust.  However, \cite{zoghbi2019} infer a black hole mass of $M/f_{X} = 4.3^{+5.2}_{-2.6} \times 10^{6}~M_{\odot}$, about four times lower than our value.  This difference is entirely attributable to the different line velocity widths that were utilized in the two analyses.  
\cite{zoghbi2019} used Chandra/HETG spectra to estimate a line width of $\sigma = 55~{\rm eV} = 2.6\times 10^{3}~{\rm km}~{\rm s}^{-1}$ \citep{miller2018}.  The resolution of the Chandra/HETG is an order of magnitude coarser than Resolve (45~eV versus 4.5~eV), and the inability to separate torus and BLR components at HETG resolution likely biased the BLR component width to low values.  The value that we have measured is roughly twice larger than that adopted by \cite{zoghbi2019}, fully accounting for the difference in $M/f_{X}$.   

The higher BLR line component width that we have measured,  $\sigma_{BLR,mean} = 5.36\pm 0.48\times 10^{3}~{\rm km}~{\rm s}^{-1}$,  makes basic physical sense.  The Fe~K$_{\alpha}$ lag is only $\tau = 3.5^{+2.8}_{-1.7}$~days, shorter than a typical H$\beta$ delay of $\tau = 5.46^{+0.87}_{-0.78}$ days \citep{bentz2022}.  A typical H$\beta$ line width in NGC~4151 is $\sigma = 1.94\pm 0.02\times 10^{3}~{\rm km}~{\rm s}^{-1}$ \citep{derosa2018}.  Given that the Fe~K$_{\alpha}$ lag is shorter than the H$\beta$ lag, it should originate at a smaller radius with a higher Keplerian velocity, and should have a higher velocity width.  In practice, differences between line velocity widths are only partly down to differences in radii; the relative geometry in the line emitting regions, how the velocities are projected into the line of sight, and gas temperatures and compositions also matter.

Our best value of the Fe~K$\alpha$ lag, $\tau = 3.5^{+2.8}_{-1.7}$~days, implies a radius of $r = 3.6^{+3.0}_{-1.7}\times 10^{3}~GM/c^{2}$ for a mass of $M_{BH} = 1.7\times 10^{7}~GM/c^{2}$ \citep{bentz2022}.  
This is formally consistent with the radii obtained in direct spectral fits to the two early XRISM/Resolve spectra ($r = 3.1^{+1.4}_{-1.2}\times 10^{3}~GM/c^{2}$ and $r = 2.8^{+0.5}_{-0.7}\times 10^{3}~GM/c^{2}$; \citealt{miller2024}).  It is also consistent with radii obtained in direct fits to the time-averaged Resolve spectrum of NGC~4151 ($r = 1.9^{+1.3}_{-0.5}\times 10^{3}~GM/c^{2}$; \citealt{miller2026}).  This suggests that spectroscopic decompositions based on Keplerian line broadening capture the key physics, despite lingering uncertainties regarding optical depths, inclinations, and opening angles.

Our results suggest that the width of the BLR component may drive changes in the flux of the Fe~K$\alpha$ line, which is nominally consistent with ``anti-breathing ''  (line width increasing in response to flux increases).  However, this phenomenon is more typically observed between monitoring campaigns, and we have only made an initial study of just nine XRISM observations; more observations are required.  Nominally, the observed variations in line width may indicate that the Fe~K$\alpha$ line is produced in a wind that responds to the central engine.  If radiation pressure is partly responsible for launching BLR winds, then ionizing flux enhancements may first stimulate stronger winds at smaller radii, potentially increasing effective line widths.  A wind scenario may be consistent with XRISM fits that find a potential role for dust in the Fe~K$\alpha$ line in NGC~4151 (\citealt{miller2024}, \citealt{miller2026}), and models of the BLR that rely on radiation pressure on dust to launch a large-scale wind \citep{czerny2015}.

Our analysis has made a number of simplifying assumptions, and might best be regarded as a pilot study.  To ensure that line width that is really due to atomic structure and scattering is not falsely attributed to velocity broadening, we modeled the narrow Fe~K$_{\alpha}$ emission line components with XCLUMPY \citep{tanimoto2019}. However, we found it necessary to fix some line parameters to values utilized in time-averaged fits \citep{miller2026}.  Our fits to each XRISM observation neglected wind absorption that is prominent between 6.5--7.0~keV.  This reflects the need for a simple, unbiased spectral model, and related work suggests that the strongest wind absorption lines are distant and not highly variable (\citealt{xiang2025}).  Our model also neglected the reflected continuum associated with the line components.  Fully self-consistent fits indicate that the reflected continuum from the BLR is $\sim$3\% of the local continuum, and it is not yet clear if the gas is optically thick or thin (see \citealt{miller2024}, \citealt{miller2026}).  This omission preserves simplicity and reproducibility at the potential expense of reduced error bars on the responding flux and smaller lag errors.  We also neglected relativistic reflection, but this is unlikely to strongly bias our results since the light travel time to the BLR is about 1000 times longer than light travel times in the inner flow.  Moreover, it is not clear if relativistic reflection is evident in NGC~4151 (e.g., \citealt{zoghbi2019}; also see \citealt{xiang2025}, \citealt{miller2026}).

The most important shortcoming in our work is likely the brevity of the XRISM monitoring.  It is likely impossible to monitor an AGN at a higher cadence with XRISM -- this would essentially amount to a continuous stare -- but a future program should hold the same cadence for 36 or 54 days (or longer), rather than 18.  This would better establish the BLR lag, and begin to probe lags from the torus.

NGC~4151 is the brightest Seyfert-1 AGN in X-rays, at energies above ${\rm E} \simeq 4$~keV.  This fact, and likely mass values that lead to favorable lag times, are at the heart of why BLR reverberation in the Fe~K$_{\alpha}$ line was first detected in this source.  A few other sources have combinations of flux and (likely) mass values that are nearly as favorable.  XRISM observations of NGC~3783 also clearly reveal BLR and torus components in its Fe~K$_{\alpha}$ emission line profile (\citealt{mehdipour2025}, \citealt{li2026}).  A preliminary lag search in XRISM observations of NGC~3783 also reveals a lag of $\tau \sim$2--3~days (Zhao et al., 2026, in preparation).  CCD spectra of NGC~3516 also reveal evidence of reverberation in the Fe~K$_{\alpha}$ line \citep{noda2023} in this AGN, making it a prime candidate for additional study with XRISM.  Usefully lower on the black hole mass scale, long exposures of NGC~4051 (rather than monitoring) may be able to reveal reverberation in its narrow Fe~K$_{\alpha}$ emission line, though this may be complicated by its status as a ``narrow-line'' Seyfert-1 AGN (see \citealt{reeves2026} for the XRISM spectrum of NGC~4051; also see \citealt{gallo2018}).  
Resolve spectra appear to detect BLR components in the narrow Fe~K$_{\alpha}$ line flux in Seyfert-2 AGN and some radio galaxies where the BLR is not visible in optical, or only tenuously inferred in polarized optical light.  The most compelling cases include Cen A \citep{bogensberger2025} and NGC~4388 \citep{fujiwara2026}.  In these and other sources that are inaccessible to optical spectroscopy, XRISM reverberation monitoring may be able to reveal the BLR, and eventually measure masses where optical studies fail (also see \citealt{minezaki2015}).

Future reverberation studies with XRISM need not be confined to the BLR.  In Table 1, it is clear that the torus component of the narrow Fe~K$_{\alpha}$ line in NGC~4151 also varies, but at a lower level than the BLR component.  More sophisticated fits to Resolve spectra of NGC~4151 suggest that the innermost face of the torus is only 2--3 times more distant from the central engine than the innermost face of the BLR (\citealt{miller2024}, \citealt{miller2026}).  Longer-term monitoring of NGC~4151 could reveal reverberation in the torus component of the Fe~K$_{\alpha}$ line.  Torus reverberation studies may be especially revealing in Seyfert-2 AGN.  Indeed, low-resolution monitoring of NGC~4388 measured an Fe~K$_{\alpha}$ lag of $\tau = 16.37^{+0.46}_{-0.37}$~days \citep{gediman2024}, potentially signaling reverberation from the innermost face of the torus.  
X-rays penetrate gas and dust, making X-ray surveys the most efficient means of discovering AGN (for a review, see, e.g., \citealt{brandt2005}).  The narrow, neutral Fe~K$_{\alpha}$ emission line is the most prominent atomic feature in AGN spectra, simply owing to the abundance of iron and its fluorescence yield (see, e.g., \citealt{gallo2023}).  This is true of unobscured Seyfert-1 AGN, and highly obscured Seyfert-2 AGN.  Future deep fields obtained with NewAthena \citep{pelle2025} are therefore expected to discover large numbers of AGN.  If XRISM stares and reverberation monitoring studies of local, bright AGN reveal Fe~K$_{\alpha}$ lags and black hole masses that correlate well with optical results, XRISM may deliver NewAthena the ability to constrain black hole masses in single-epoch fits.  In short, a black hole mass may be possible for every AGN bright enough to register a spectrum that can be separated into BLR and torus components.

\hspace{0.2in}
We thank the anonymous referee for a thoughtful and thorough review of this work that led to many improvements.  We thank the Swift and XRISM scheduling and operations teams for making this program possible.  JMM acknowledges helpful conversations with Jelle de Plaa.  JMM acknowledges support from NASA through the XRISM Guest Observer program.  AZ is supported by NASA under award number 80GSFC24M0006.

\bibliography{main}{}
\bibliographystyle{aasjournalv7}

\begin{figure*}
    \centering
    \includegraphics[width=0.9\textwidth]{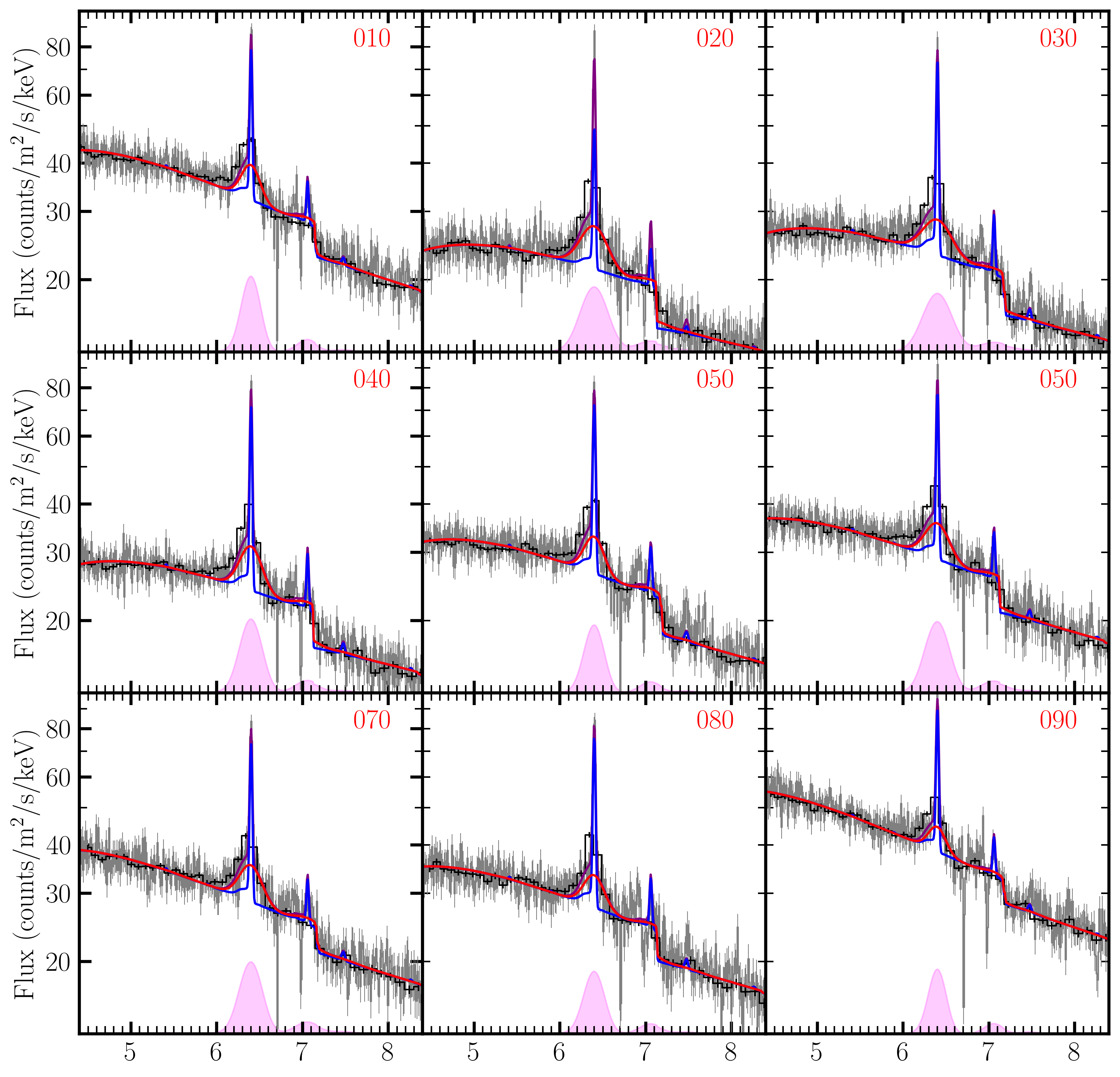}
    \caption{The nine XRISM spectra of NGC~4151, centered on the narrow Fe~K$-{\alpha}$ emission line.  The data are binned for visual clarity.  The Resolve spectrum is shown in gray, and the Xtend spectrum in black.  For clarity, only the Resolve model components are shown (purple: total model, blue: torus line, red: BLR line).  The shaded magenta region depicts the flux of the BLR line without the continuum.  Note that the BLR line varies in peak strength and width, and the total line flux depends on both parameters.}
    \label{fig:broadspectra}
\end{figure*}

\begin{figure*}
    \centering
    \includegraphics[width=0.9\textwidth]{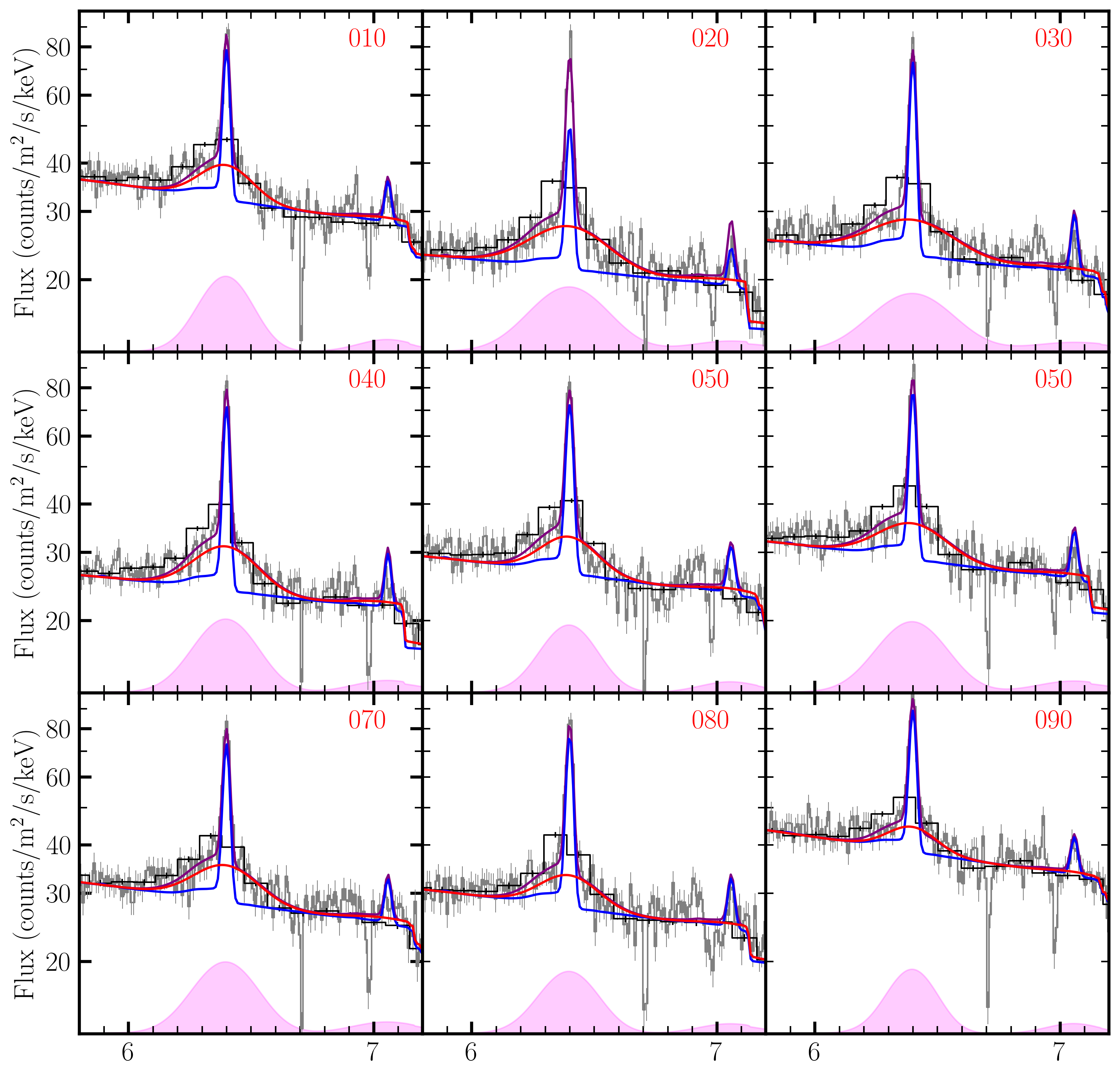}
    \caption{Same as Figure 1, but the spectra are shown on a narrower band to illustrate additional details of the Fe~K$_{\alpha}$ emission line components.}
    \label{fig:spectra}
\end{figure*}

\begin{table}[t!]
\caption{Key Spectral Parameters}
\begin{scriptsize}
\begin{center}
\begin{tabular}{llllllllll}
Short ObsID & 010 & 020 & 030 & 040 & 050 & 060 & 070 & 080 & 090 \\
\tableline
MJD (midpoint) & 60657.38 & 60659.53 & 60661.81 & 60664.11 & 60666.39 & 60668.52 & 60670.97 & 60672.73 & 60674.49 \\ 
\tableline
${\rm F}_{\rm BLR}~(10^{-12}~{\rm erg}~{\rm cm}^{-2}~{\rm s}^{-1})$ & $2.60^{+0.14}_{-0.14}$ & $3.10^{+0.15}_{-0.15}$ & $2.91^{+0.16}_{-0.16}$ & $2.93^{+0.12}_{-0.12}$ & $2.39^{+0.13}_{-0.13}$ & $3.09^{+0.16}_{-0.16}$ & $2.92^{+0.16}_{-0.16}$ & $2.32^{+0.16}_{-0.16}$ & $2.18^{+0.18}_{-0.18}$ \\

$\sigma_{\rm BLR}~(10^{3}~{\rm km}~{\rm s}^{-1})$  & $4.8^{+0.4}_{-0.4}$ & $6.0^{+0.5}_{-0.5}$ & $6.4^{+0.6}_{-0.6}$ & $5.5^{+0.5}_{-0.4}$ & $5.2^{+0.5}_{-0.4}$ & $5.8^{+0.6}_{-0.6}$ & $4.9^{0.5}_{-0.5}$ & $5.1^{+0.7}_{-0.6}$ & $4.5^{+0.6}_{-0.6}$\\

\tableline

${\rm F}_{\rm TOR}~(10^{-12}~{\rm erg}~{\rm cm}^{-2}~{\rm s}^{-1})$ & $2.41^{+0.09}_{-0.09}$ & $2.60^{+0.08}_{-0.08}$ & $2.61^{+0.08}_{-0.08}$ & $2.51^{+0.07}_{-0.07}$ & $2.66^{+0.08}_{-0.08}$ & $2.67^{+0.10}_{-0.10}$ & $2.43^{+0.09}_{-0.09}$ & $2.74^{+0.10}_{-0.10}$ & $2.84^{+0.11}_{-0.11}$ \\

$\sigma_{\rm TOR}~(10^{2}~{\rm km}~{\rm s}^{-1})$ & $5.11^{+0.46}_{-0.42}$ & $5.0^{+0.4}_{-0.4}$ & $4.8^{+0.3}_{-0.3}$ & $5.1^{+0.3}_{-0.3}$ & $6.0^{+0.4}_{-0.4}$ & $5.4^{+0.4}_{-0.4}$ & $5.0^{+0.4}_{-0.4}$ & $5.4^{+0.4}_{-0.4}$ & $5.6^{+0.4}_{-0.4}$ \\

\tableline

$\Gamma$ & $2.01^{+0.02}_{-0.02}$ & $1.96^{+0.01}_{-0.01}$ & $1.95^{+0.01}_{-0.01}$ & $1.97^{+0.01}_{-0.01}$ & $2.02^{+0.01}_{-0.01}$ & $2.00^{+0.01}_{-0.01}$ & $1.93^{+0.01}_{-0.01}$ & $1.87^{+0.01}_{-0.01}$ & $1.94^{+0.01}_{-0.01}$\\

${\rm Norm.}~(10^{51}~{\rm ph}~{\rm s}^{-1}~{\rm keV}^{-1})$ & $4.61^{+0.17}_{-0.18}$ & $2.89^{+0.01}_{-0.01}$ & $3.36^{+0.01}_{-0.01}$ &  $3.23^{+0.01}_{-0.01}$ & $4.15^{+0.01}_{-0.01}$ & $3.92^{+0.01}_{-0.01}$ & $3.50^{+0.01}_{-0.01}$ & $3.08^{+0.01}_{-0.01}$ & $4.49^{+0.01}_{-0.01}$ \\

${\rm N}_{\rm H}~(10^{23}~{\rm cm}^{-2})$ & $1.55^{+0.02}_{-0.02}$ & $2.07^{+0.06}_{-0.06}$ & $2.08^{+0.05}_{-0.05}$ & $1.90^{+0.05}_{-0.05}$ & $1.91^{+0.05}_{-0.05}$ & $1.69^{+0.05}_{-0.05}$ & $1.49^{+0.06}_{-0.06}$ & $1.51^{+0.04}_{-0.04}$ & $1.24^{+0.04}_{-0.04}$\\

\tableline

$C/\nu$ & 872.6/450 & 684.4/437 & 790.9/444 & 771.4/445 & 851.0/443 & 716.6/442 & 677.4/440 & 634.3/437 & 723.6/451  \\
\tableline
\end{tabular}
\end{center} 
\tablecomments{Spectral fitting parameters from joint fits to the Resolve and Xtend spectra of NGC~4151 in the 4.4--8.4~keV band (centered on Fe~K$_{\alpha}$ at 6.40~keV).  The full observation identification numbers (ObsIDs) are, e.g., 201076010.  The MJD column lists the midpoint of each XRISM observation.  The other parameters listed include the flux and width of the BLR line component, the flux and width of the torus line component, the index, normalization, and flux in the power-law continuum, the column density of the cool internal absorber, and the Cash statistic divided by the degrees of freedom.  Fluxes are quoted over the ionizing band, 0.0136--13.6~keV.  Note that the fits are not formally acceptable because complex absorption above the emission line complex is not modeled.}
\end{scriptsize}
\end{table}

\begin{figure*}
    \centering
    \includegraphics[width=0.7\textwidth]{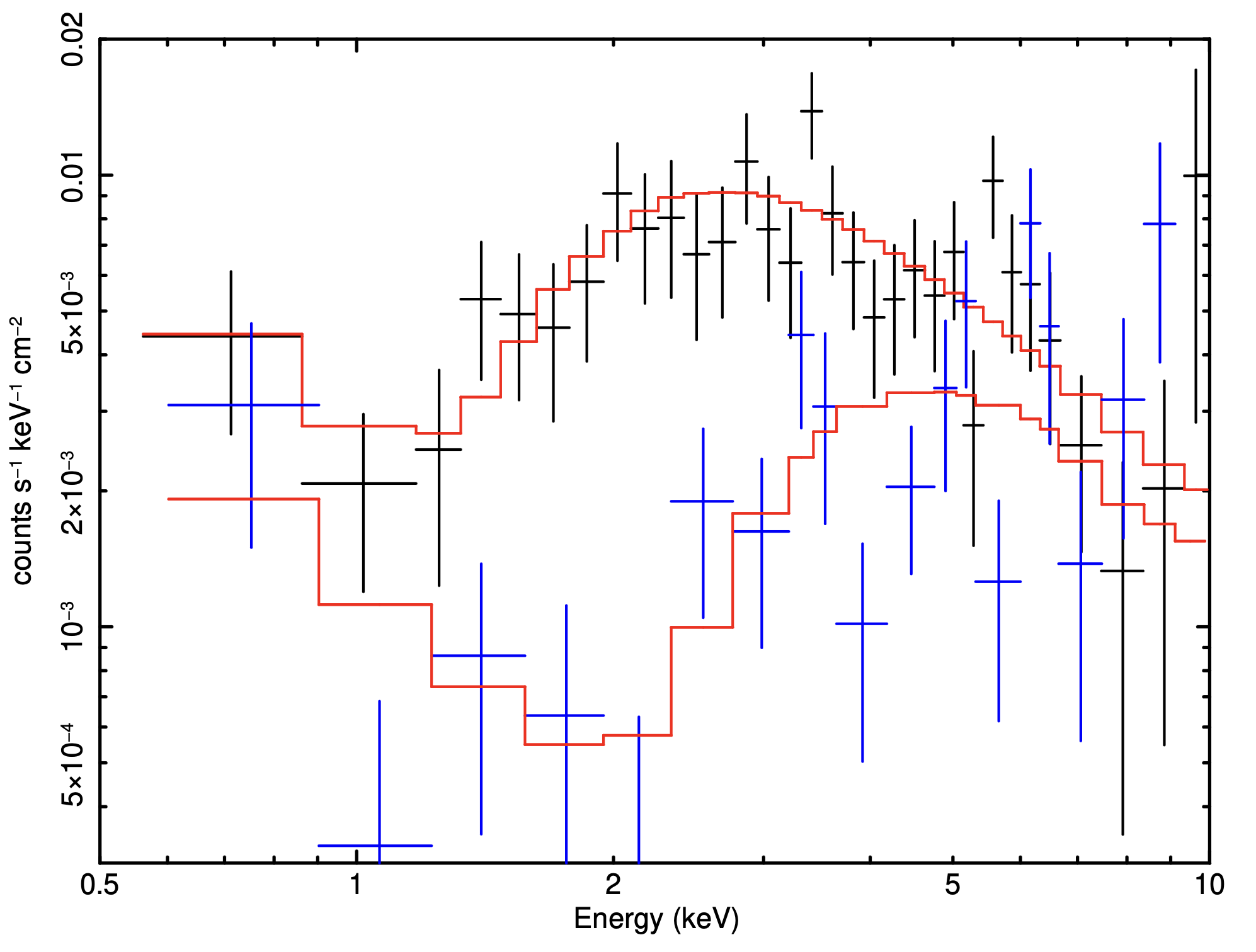}
    \caption{Swift/XRT spectra of NGC~4151 obtained as part of our two-month continuum monitoring campaign.  The spectra shown here represent the extrema of the variable internal column(s) that shape the spectrum of NGC~4151.  In this analysis, changes in the unabsorbed flux are estimated by modeling the data using neutral partial-covering absorption, and a simple power-law continuum.  This simple treatment is a proxy for complex ionized absorption.  The most highly obscured spectrum in the sequence (shown in blue) was obtained on MJD 60661.10, and the least obscured was obtained on MJD 60704.39 (shown in black).}
    \label{fig:xrtspec}
\end{figure*}

\begin{figure*}
    \centering
    \includegraphics[width=0.7\textwidth]{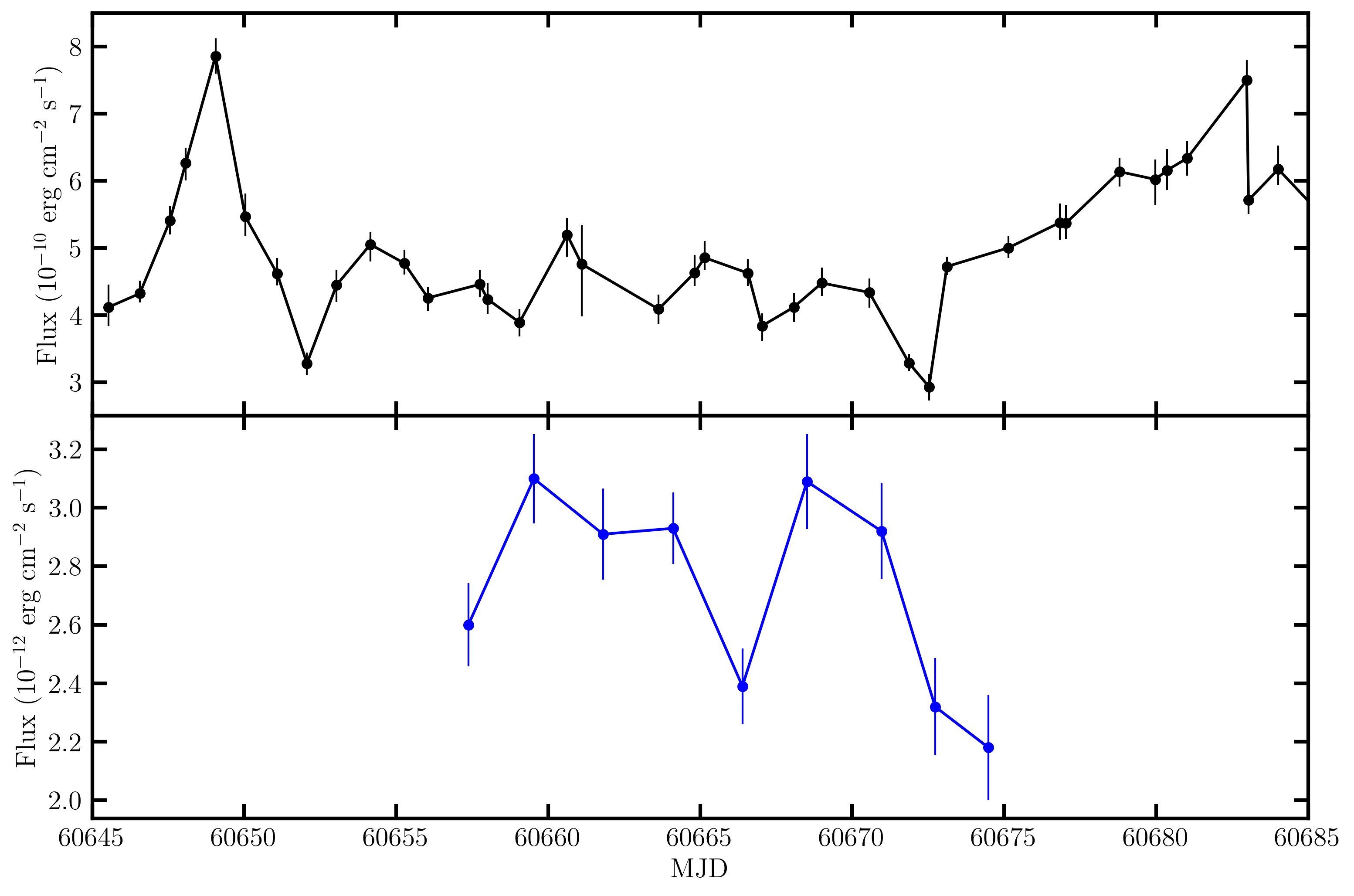}
    \caption{Upper panel: the unabsorbed X-ray flux in the full XRISM/Resolve band (2.2-17.4~keV), as measured via Swift/XRT monitoring observations.  Lower panel: the flux of the BLR component of the Fe~K$_{\alpha}$ emission lines, as measured via joint fits to Resolve and Xtend spectra.}
    \label{fig:amps}
\end{figure*}

\begin{figure*}
    \centering
    \includegraphics[width=0.49\textwidth]{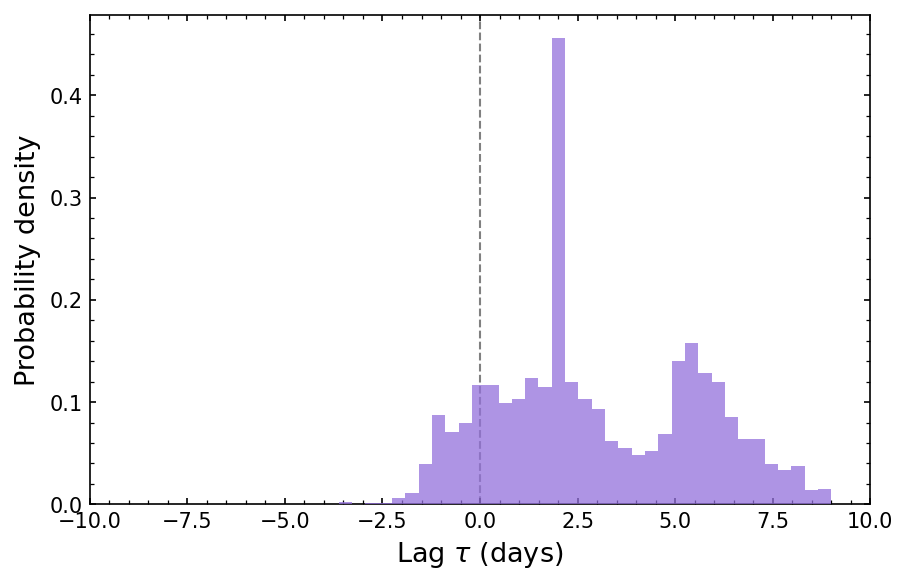}
    \includegraphics[width=0.49\textwidth]{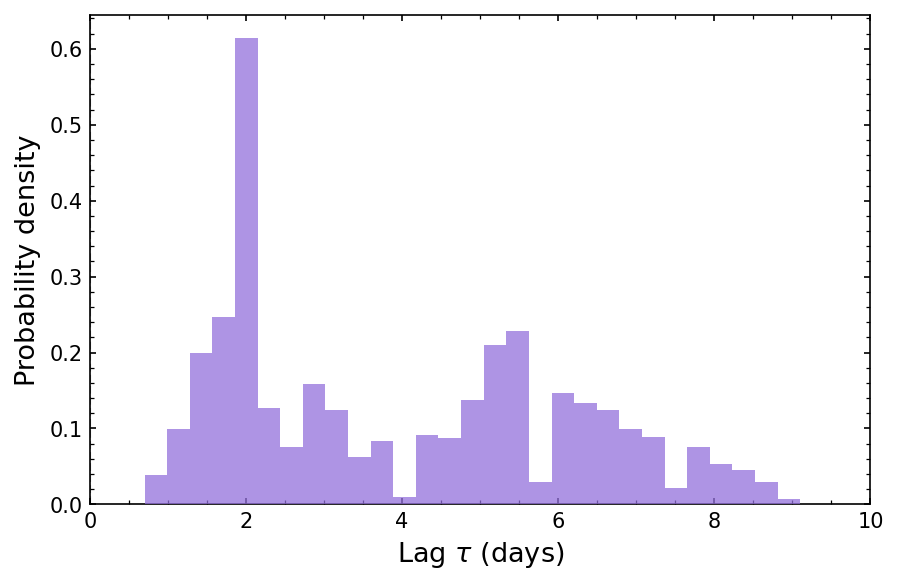}
    \caption{The cross correlation centroid lag probability distributions that result from running pyCCF with 10,000 Monte Carlo simulations over a range of [-5,10] days (left) and [0,10] days (right).  The distribution shown at left verifies that there are no spurious peaks below zero lag.  The distribution at right indicates that the BLR component of the Fe~K$_{\alpha}$ line lags the driving continuum by $\tau = 3.5^{+2.8}_{-1.7}$~days.  Fully consistent results are obtained by running Javelin instead of pyCCF.}
    \label{fig:spectra}
\end{figure*}

\begin{figure*}
    \centering
    \includegraphics[width=0.7\textwidth]{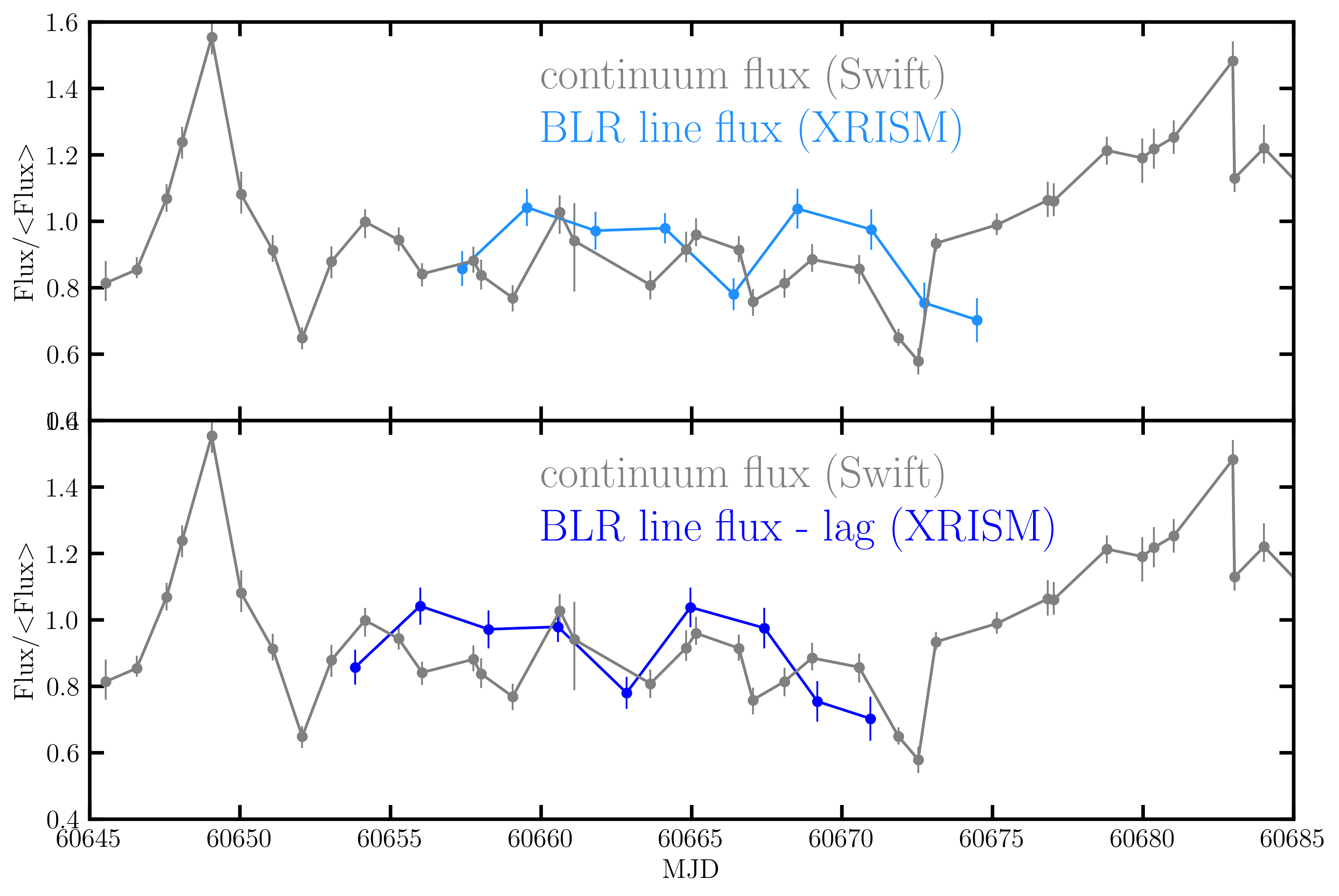}
    \caption{Continuum and BLR line flux trends, relative to local mean values.  The XRISM data are vertically shifted to better enable comparisons with local trends in the continuum flux.  When shifted by the best-fit lag, the BLR line flux more closely matches the continuum flux trend.}
    \label{fig:amps}
\end{figure*}



\begin{table*}[h!]
\centering
\begin{tabular}{c c@{\hspace{1.5cm}} c c}
MJD & Flux & MJD & Flux \\
\hline
60645.53 & $4.12^{+0.34}_{-0.28}$ & 60675.14 & $5.00^{+0.17}_{-0.15}$ \\
60646.57 & $4.32^{+0.19}_{-0.13}$ & 60676.84 & $5.38^{+0.28}_{-0.26}$ \\
60647.56 & $5.41^{+0.21}_{-0.21}$ & 60677.03 & $5.37^{+0.27}_{-0.23}$ \\
60648.08 & $6.27^{+0.23}_{-0.26}$ & 60678.80 & $6.14^{+0.21}_{-0.22}$ \\
60649.06 & $7.86^{+0.26}_{-0.26}$ & 60679.98 & $6.02^{+0.29}_{-0.38}$ \\
60650.03 & $5.47^{+0.34}_{-0.30}$ & 60680.36 & $6.16^{+0.31}_{-0.30}$ \\
60651.08 & $4.62^{+0.23}_{-0.18}$ & 60681.02 & $6.34^{+0.26}_{-0.26}$ \\
60652.07 & $3.28^{+0.16}_{-0.17}$ & 60682.98 & $7.50^{+0.30}_{-0.30}$ \\
60653.04 & $4.45^{+0.23}_{-0.26}$ & 60683.04 & $5.71^{+0.18}_{-0.21}$ \\
60654.15 & $5.05^{+0.19}_{-0.25}$ & 60684.02 & $6.17^{+0.35}_{-0.24}$ \\
60655.27 & $4.77^{+0.19}_{-0.17}$ & 60685.33 & $5.54^{+0.18}_{-0.22}$ \\
60656.05 & $4.26^{+0.17}_{-0.19}$ & 60686.04 & $5.55^{+0.21}_{-0.24}$ \\
60657.75 & $4.46^{+0.21}_{-0.18}$ & 60687.87 & $6.39^{+0.25}_{-0.27}$ \\
60658.01 & $4.23^{+0.24}_{-0.22}$ & 60688.14 & $4.39^{+0.28}_{-0.22}$ \\
60659.06 & $3.89^{+0.20}_{-0.21}$ & 60689.05 & $4.38^{+0.15}_{-0.20}$ \\
60660.62 & $5.19^{+0.25}_{-0.33}$ & 60690.55 & $5.22^{+0.30}_{-0.30}$ \\
60661.10 & $4.76^{+0.57}_{-0.78}$ & 60691.60 & $5.01^{+0.28}_{-0.32}$ \\
60663.63 & $4.09^{+0.22}_{-0.22}$ & 60692.72 & $6.40^{+0.37}_{-0.29}$ \\
60664.82 & $4.63^{+0.27}_{-0.20}$ & 60693.43 & $5.18^{+0.38}_{-0.42}$ \\
60665.15 & $4.85^{+0.25}_{-0.18}$ & 60694.87 & $5.38^{+0.47}_{-0.32}$ \\
60666.57 & $4.62^{+0.21}_{-0.19}$ & 60695.06 & $4.79^{+0.28}_{-0.33}$ \\
60667.03 & $3.84^{+0.19}_{-0.22}$ & 60696.56 & $4.12^{+0.25}_{-0.32}$ \\
60668.08 & $4.12^{+0.21}_{-0.22}$ & 60697.94 & $6.09^{+0.37}_{-0.38}$ \\
60669.00 & $4.48^{+0.23}_{-0.19}$ & 60699.62 & $6.19^{+0.29}_{-0.27}$ \\
60670.56 & $4.34^{+0.21}_{-0.23}$ & 60700.54 & $3.69^{+0.23}_{-0.18}$ \\
60671.88 & $3.28^{+0.14}_{-0.12}$ & 60701.19 & $4.62^{+0.28}_{-0.27}$ \\
60672.53 & $2.93^{+0.19}_{-0.20}$ & 60703.54 & $6.80^{+0.36}_{-0.57}$ \\
60673.12 & $4.72^{+0.15}_{-0.13}$ & 60704.40 & $5.45^{+0.34}_{-0.30}$ \\
\hline
\end{tabular}
\caption{Swift/XRT measurements of the continuum flux, based on absorbed power-law fits in the 0.3--10.0~keV band and projected into the 2.4--17.4~keV band to match prior XRISM studies \citep{miller2024,xiang2025,miller2026}.
Fluxes are given in units of
$10^{-10}\,\mathrm{erg\,cm^{-2}\,s^{-1}}$.}
\label{tab:swiftfluxes}
\end{table*}

\end{document}